\documentclass{article}
\usepackage[utf8]{inputenc}

\title{Investigating The Hubble Tension Through Hubble Parameter Data}
\author{thakurr58 }
\date{January 2020}

\usepackage[utf8]{inputenc}
\usepackage{amsmath}
\usepackage{amsfonts}
\usepackage{amssymb}
\usepackage[english]{babel}

\usepackage{color}
\usepackage{epsfig}
\usepackage{cite}
\usepackage{graphicx}%
\usepackage{dcolumn}%
\usepackage{bm}%

\begin{document}
\begin{center}
{\large Investigating The Hubble Tension Through Hubble Parameter Data} \\
\vspace*{0.5cm}
Rahul Kumar Thakur\textsuperscript{c,}\footnote{thakurr58@gmail.com}, Shashikant Gupta\textsuperscript{b,}\footnote{shashikantgupta.astro@gmail.com}, Rahul Nigam\textsuperscript{a,}\footnote{rahul.nigam@hyderabad.bits-pilani.ac.in}, PK Thiruvikraman \textsuperscript{a,}\footnote{	
thiru@hyderabad.bits-pilani.ac.in}\\
\vspace*{0.5cm}
$^a$ BITS Pilani, Hyderabad Campus, Hyderabad, India.\\
$^b$ G D Goenka University,\\ Gurgaon, Haryana, 122103, India.\\
$^c$ Avantika University, Ujjain.\\
\vspace*{0.1cm}
\end{center}

\begin {abstract}
%\textbf{Context/Background:} 
The Hubble constant ($H_0$), which represents the expansion rate of the Universe, is one of the most important cosmological parameters. The recent measurements of $H_0$ using the distance ladder methods such as Type Ia Supernovae (SNe Ia) are significantly greater than the CMB measurements by Planck. The difference points to a crisis in the standard model of cosmology termed as Hubble tension.
 In this work we compare different cosmological models, determine the Hubble constant and comment on the Hubble tension using the data from differential ages of galaxies. The data we use is free from the systematic effects as the absolute age estimation of the galaxies is not needed. We have used the Bayesian approach along with the commonly used maximum likelihood method to estimate $H_0$ and have calculated the AIC scores to compare the different cosmological models.The non-flat cosmological model provides a higher value for matter density as well as the Hubble constant compared to the flat $\Lambda$CDM model. The AIC score is smaller for the flat $\Lambda$CDM cosmology compared to the non-flat model indicating the flat model a better choice. The best-fit value of $H_0$ for both these models are $68.7\pm3.1$ km/s/Mpc and $72.2\pm4$ km/s/Mpc, respectively. Our results are consistent with the CCHP measurements. However, flat model result does not agree with the SH0ES result, while the non-flat result is inconsistent with the Planck value. 

\end{abstract}

%{\it {\bf Keywords:} Cosmology, Galaxies, CMBR, Supernovae, Hubble constant}
\newpage

\section{Introduction}
\label{sec:intro}
The linear relation between the distance to various galaxies from us and their recessional velocities was the first evidence for the state of expansion of the Universe \cite{hubble1929}. The slope of the graph, also known as the Hubble constant, measures the expansion rate of the universe. Additionally, observations of the type Ia supernovae (SNe) shows that the expansion is accelerated \cite{riess1998observational,perlmutter1998discovery}. The $\Lambda CDM$ model \cite{astier2012observational} is the simplest cosmological model which provide a good fit for available cosmological data. 

The Hubble constant ($H_0$) is one of the most important parameter in modern cosmology; and along with other cosmological parameters it sets the age, size and shape of the universe. Determining the accurate value of $H_0$ is a challenging task  
for cosmologists since last few decades. Measuring the value within 10\% accuracy has been one of the key projects of the Hubble Space Telescope. The current estimate of the key project is $H_0 = 72\pm 8$ km/S/Mpc \cite{freedman2001final}. Lately some excellent progress has been made towards measuring the Hubble constant as a number of different methods of measuring distances have been developed and refined. Supernovae, $H_0$, Equation of State of Dark energy (SH0ES) is among the most precise measurements of type Ia SNe distances for the above purpose. 
From the SH0ES program \cite{riess20162} a value of $H_0 = 73.24\pm 1.74$ km/S/Mpc was obtained. On the other hand observations of Cosmic Microwave Background (CMB) anisotropies can also provide a global value of $H_0$ when $\Lambda CDM$  cosmology is applied to it. Coincidentally, these two measurements of the Hubble constant disagree at more than 3 $\sigma$ level. The discrepancy is termed as  “Hubble tension” \cite{dainotti2021hubble}. The conflict is alarming and it  possibly indicates a new physics beyond the standard $\Lambda$CDM cosmological model \cite{freedman2017cosmology,feeney2018clarifying,de2020addressing,vagnozzi2020new,vagnozzi2021consistency}  
Recently, \cite{freedman2019carnegie} have calibrated type Ia SNe using tip of red giant branch (TRGB) stars. Their value of $H_0 = 69.8\pm 0.8 \pm 1.1$ km/S/Mpc is smaller than \cite{riess20162} leading to a reduction in the discrepancy level. However, the tension between the local and global value of $H_0$ has not disappeared and requires attention of the researchers.

We plan to analyse the Hubble parameter data sets using flat and non-flat $\Lambda$CDM model by using Bayesian analysis to test if the Hubble tension is real. This paper is organised as follows: The data and method of our analysis are presented in section \ref{sec:method}. The results and conclusions have been presented in section \ref{sec:results} and \ref{sec:conclusion}.

%%%%%%%%%%%%%%%%%%%%%%%%%%%%%%%%%%%%%%
\section{Methodology and Data}
\label{sec:method}
We begin with the maximum likelihood method which is a common approach to estimate best-fit parameters for a model. One can define the likelihood in terms of $\chi^2$ as follows: 
\begin{equation}
P(D|M) \propto \exp{(-\chi^2/2)} \,  .
\label{eq:likeli}
\end{equation}
Where likelihood, $P(D|M)$, is the probability of obtaining the data assuming that the given cosmological model $M$ is correct. In the present analysis we have considered the flat and non-flat $\Lambda$CDM cosmological models. One can maximize the likelihood or minimize $\chi^2$ with respect to the model parameters to obtain the best-fit. $\chi^2$ is define for the above cosmological models as 
\begin{equation}
%    \chi^2(H_{0},\Omega_{M}) = \Sigma_{i=1}^{31} \left[\frac{H^{th}(z_{i},p)-H_{i}^{obs}}{\sigma_{H,i}} \right]^2
\chi^2(a_j) = \sum_{i} \left[\frac{H^{th}(z_{i},a_j)-H_{i}^{obs}}{\sigma_{H_i}} \right]^2
    \label{eq:chi-1}
\end{equation}
where, the free parameters of our model, $a_j$, are $\Omega_M$ and $H_0$ for flat while $\Omega_k$, $\Omega_M$ and $H_0$ for non-flat cosmology. $H^{th}$ and $H^{obs}$ denote the theoretical and observed value of the Hubble parameter while $\sigma_{H_i}$  stands for the standard error in $H^{obs}$. The  Hubble parameter $H^{th}$ for spatially flat $\Lambda$CDM model is
\begin{equation}
    H^{th} = H_{0} \sqrt{\Omega_{M} (1+z)^3 + 1-\Omega_{M}} \, ,
    \label{eq:h-th}
\end{equation}
where $\Omega_{M}$ is the present value of the density parameter. In the non-flat $\Lambda$CDM model the expansion rate function is given by
\begin{equation}
    H^{th} = H_{0} \sqrt{\Omega_{M} (1+z)^3 + \Omega_{k} (1+z)^2 + 1-\Omega_{M} - \Omega_{k}} \, ,
    \label{eq:h-nonflatth}
\end{equation}
where ${\Omega_{k}}$ is the current value of the spatial curvature energy density parameter.

%%%%%%%%%%%%%%%%%%%%%%
\subsection{Akaike Information Criterion}
The Akaike Information Criterion, popularly known as AIC, is a technique for assessing how well the data fits a specific model. It is used to compare different models to determine which one fits the data better. AIC can be computed from the likelihood, $L$, and the number of independent variables, $k$, in the following manner \cite{akaike1974new}
\begin{equation}
    AIC = -2log L + 2K \, .
    \label{eq:aic}
\end{equation}
Smaller value of AIC indicates a better fit. A difference of more than $2$ AIC units between the AIC scores of different models is considered significant. The default value of $K$ is $2$ with no independent parameters. Here we shall compare the flat $\Lambda$CDM model with $\Omega_M$ and $H_0$ as independent variables with the non-flat $\Lambda$CDM model in which $\Omega_k$ is an additional independent variable. Our cosmological models have two and three parameters, respectively, hence the values of $K$ are $4 and $5 for each model.

%%%%%%%%%%%%%%%%%%%
\subsection{The Bayesian Approach} 
We use both the maximum likelihood method as well as the Bayesian approach to estimate the best-fit values of cosmological parameters.  
The posterior probability of the parameters can be calculated using Bayes theorem
\begin{equation}
     P(M|D) \propto P(D|M)\times P(M) \, 
    \label{eq:bayes}
\end{equation}

The main criticism of the Bayesian approach arises from the prior probability which represents our state of knowledge about the model itself since it could be subjective. One should be careful while selecting the prior probability, and stringent priors should be avoided. However the Bayesian approach is useful as it allows one to calculate the direct probability of model parameters. 
%The other advantage of this approach is the marginalization over the nuisance parameters. For instance the cosmological models often require the matter density ($\Omega_M$) and the expansion rate ($H_0$) of the universe as the model parameters. Since, we are interested in the Hubble constant, we marginalize over the density parameter $\Omega_M$ through the following equation:
The other advantage of this approach is the marginalization over the undesired model parameters. For instance, $\Omega_M$ and $H_0$ are often used as the essential parameters in most of the cosmological models. Since, we are interested in the expansion rate only, we prefer marginalizing over $\Omega_M$ using the following equation:
\begin{equation}
    P({H_{0}) = \int P( \Omega_{M},H_{0})P(\Omega_{M},H_{0})d\Omega_{M}}
    \label{eq:margin}
\end{equation}
Two different types of priors have been considered in our analysis: i) uniform prior ($0\leq \Omega_M \leq1$) and ii) Gaussian priors centered around the best-fit value. We have carefully chosen the prior probability of $\Omega_M$ within a reasonable range. 

%%%%%%%%%%%%%%%%%%%%%%
\subsection{H(z) Data and the Differential Ages of Galaxies} 
The data set consist of 31 H(z) values,recently compiled by \cite{cao2022using}. The redshift range covered by the measurements is $z \leq 0.07 \leq 2.42 $. Earlier attempts of estimating $H_0$ from Hubble parameter data can be found in \cite{cao2022using}. This technique uses passively evolving early–type galaxies and does not depend on the cosmological model. 
This method can provide constraints on the cosmological parameters as it does not rely on the nature of metric between the observer and the chronometers. 
The differential approach instead of the real ages of the galaxies is the reason for the above advantage. Additionally, this technique is immune to systematic effects as the absolute age estimation of the galaxies is not required. Luminous red galaxies (LRGs) are regarded as a good candidate for this method as their photometric properties are consistent with an old passively evolving stellar population. 

%%%%%%%%%%%%%%%%%%%%%%%%%%%%%%%%%%%%%%
\begin{center}
\begin{table}
\centering
\begin{tabular}{ |c|c|c|c|c|} 
 \hline
 $\Omega_{M}$ & $H_0$ &$\chi_{\nu}^{2}$ & $AIC$ \\ 
 \hline
 0.28 & 68.8 & 0.972 & 36.21\\
 \hline
\end{tabular}
\caption{Best-fit value of parameters for a flat $\Lambda CDM$ cosmology from H(z) data by minimising $\chi^2$.}
\label{Table:chisq}
\end{table}
\end{center}
%%%%%%%%%%%%%%%%%%%%%%%%%%%%%%
\begin{center}
\begin{table}
\centering
\begin{tabular}{ |c|c|c|c|c|c|} 
 \hline
 $\Omega_{M}$ & $H_0$ & $\Omega_{k}$ & $\chi_{\nu}^{2}$ & $AIC$ \\ 
 \hline
 0.45 & 73.1 & -0.53 & 0.973 & 37.25\\
 \hline
\end{tabular}
\caption{Best-fit values of Cosmological parameters by minimising $\chi^2$  for non- flat $\Lambda$CDM model.}
\label{Table:non-flatchisq}
\end{table}
\end{center}
%%%%%%%%%%%%%%%%%%%%%%   
 \begin{center}
\begin{table}
\centering
\begin{tabular}{|c|c|c|c|} 
 \hline
 Probe & Model & $H_0$ & $\sigma$  \\
 \hline
% Diff. Ages & Flat $\Lambda CDM$ with Uniform Prior & 68.7 & 3.1 \\
 Diff. Ages & Flat $\Lambda CDM$ & 68.7 & 3.1 \\
 \hline
% Diff. Ages & Non-flat $\Lambda CDM$ with Uniform Prior & 72.2 & 4 \\
  Diff. Ages & Non-flat $\Lambda CDM$ & 72.2 & 4 \\
 \hline
 Planck \cite{ade2014planck} & - & 67.8 & 0.90 \\
 \hline
 SH0ES \cite{riess20162} & - & 73.24 & 1.74 \\
 \hline
 CCHP \cite{freedman2019carnegie} & - & 69.8 & 0.8 \\
 \hline
\end{tabular}
\caption{$H_{0}$ bestfit values after marginalization from the Hubble parameter data. Both the Gaussian and uniform priors in reasonable range provide same values of $H_0$. Other measurements from the literature are shown for comparison.}
\label{Table:margin}
\end{table}
\end{center}
%%%%%%%%%%%%%%%%%%%%%%%%%%%%%%%%%%%%%%
\vspace{-2cm}
\section{Results and Discussion}
\label{sec:results}
We first calculate the best-fit parameters from the $H(z)$ data set by minimising $\chi^{2}$ defined in Eq.~\ref{eq:chi-1}. The minimum value of $\chi^{2}$ and the best-fit cosmological parameters for both the flat and non-flat $\Lambda$CDM model are presented in Table-\ref{Table:chisq} and \ref{Table:non-flatchisq}. It is clear that $\chi_{\nu}^{2}$ is smaller than 1, indicating that the error bars probably have been overestimated. The large error bars in the data also indicate the same. A comparison of the tables show that flat $\Lambda$CDM favors lower matter density and Hubble constant compared to the non-flat $\Lambda$CDM model. We further calculate the likelihood and AIC score for both the models using Eq~\ref{eq:likeli} and \ref{eq:aic}. The AIC score for flat cosmology is smaller and hence this model should be favored.  
Now we apply the Bayesian analysis and calculate the posterior probability using Eq~\ref{eq:bayes}. Finally marginalization over the matter density, $\Omega_M$, is performed and the corresponding best-fit value of Hubble constant is calculated which is presented in table-\ref{Table:margin}. Both Gaussian as well as uniform priors have been used for the marginalization.  
The best-fit value of $H_0$ are almost same in the two cases of marginalization.
For non-flat $\Lambda$CDM cosmology, marginalisation over $\Omega_k$ and $\Omega_M$ have been performed. The final value of $H_0$ is shown in Table~\ref{Table:margin} which is again higher than the value obtained for flat cosmology. 

%%%%%%%%%%%%%%%%%%%%%%%%%%%%%%%%%%%

\begin{figure}
\centering
\includegraphics[width=10.0cm]{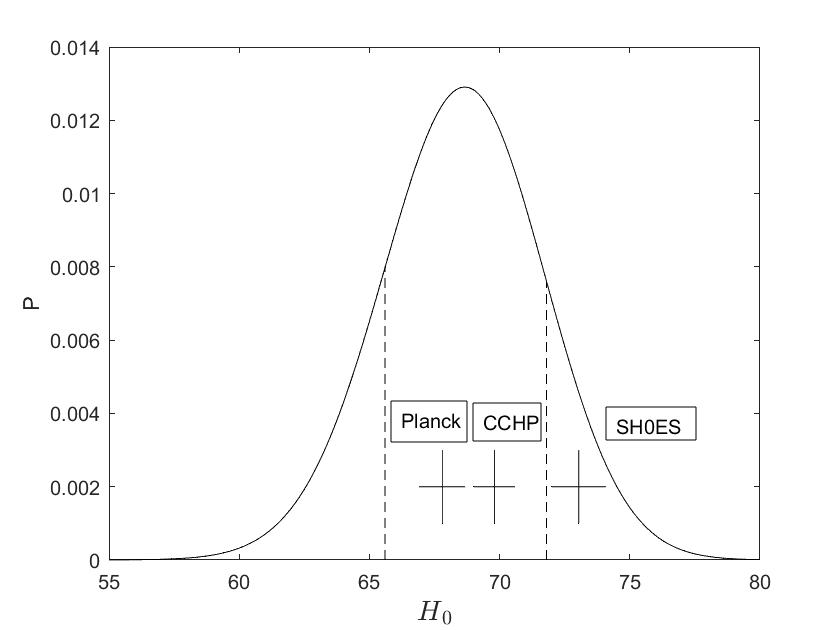}
\caption{Probability Distribution of $H_{0}$ values for flat $\Lambda$CDM model after marginalization over $\Omega_M$ for Hubble Parameter Data. Planck and CCHP values are within 1 $\sigma$ however SH0ES value is outside of the 1 $\sigma$ level.}
\label{fig:hubble}
\end{figure}

%%%%%%%%%%%%%%%%%%%%
\begin{figure}
\centering
\includegraphics[width=10.0cm]{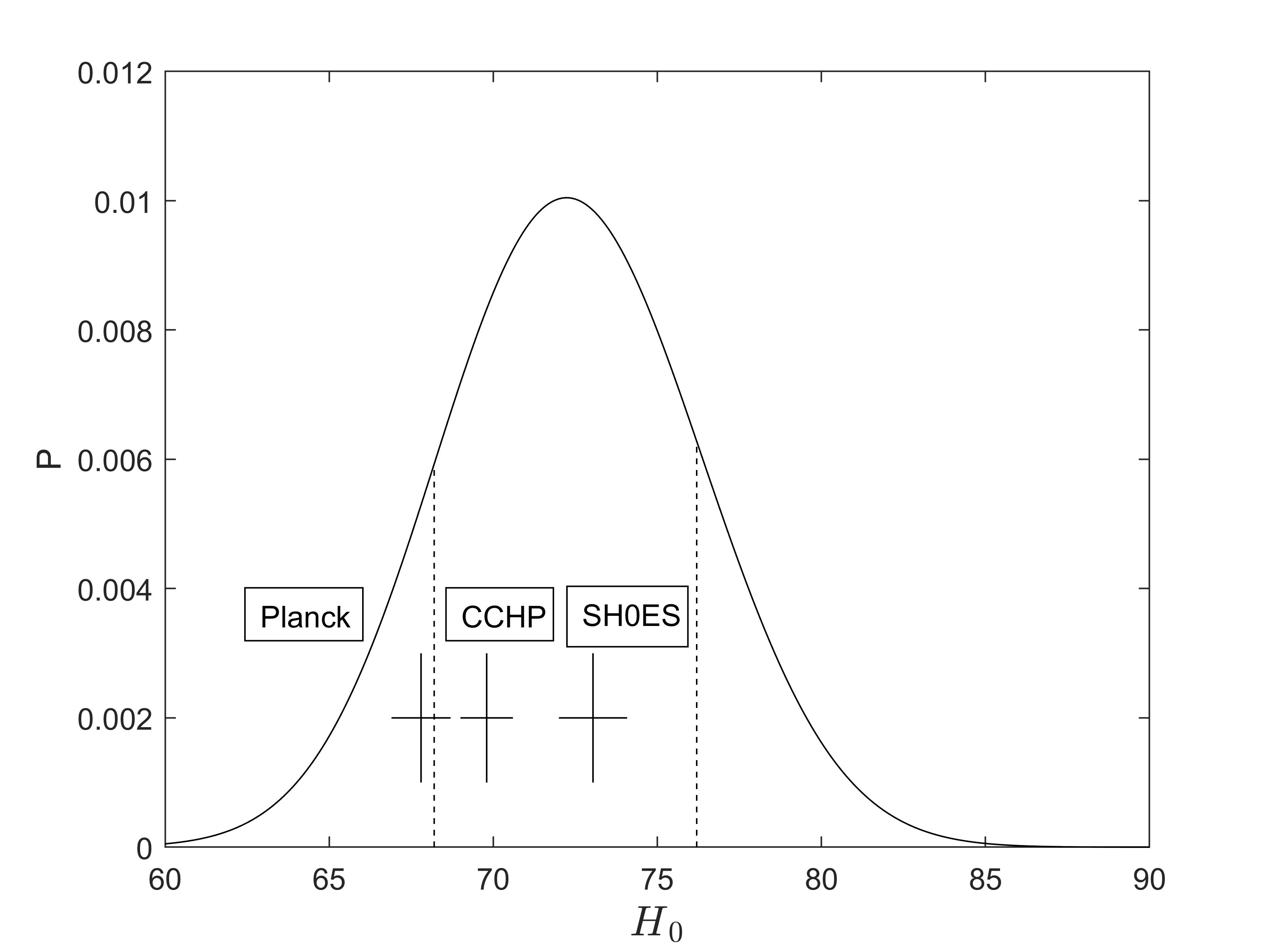}
\caption{Probability Distribution of $H_{0}$ values for non- flat $\Lambda$CDM model after marginalization over $\Omega_k$ and $\Omega_M$ for Hubble Parameter Data. Planck value lies outside of 1 $\sigma$ however CCHP and SH0ES values are within the 1 $\sigma$ level.}
\label{fig:nonflathubble}
\end{figure}

Finally, we compare the numerical value of $H_0$ for both flat and non-flat $\Lambda$CDM model obtained from the Hubble parameter data with the latest measurements of $H_0$. The posterior probability of $H_0$ for flat $\Lambda$CDM cosmology from the $H(z)$ data is plotted in figure~\ref{fig:hubble}. The best-fit value is $68.7$ and the area between the vertical dash lines corresponds to 1~$\sigma$ confidence level. For comparison, the $H_0$ values from Planck, SH0ES collaboration and Carnegie-Chicago Hubble Program (CCHP) \cite{freedman2019carnegie} have also been shown in the same graph. Planck and CCHP values are within the 1~$\sigma$ region of our result. However, SH0ES value is higher than all other values and lies outside 1~$\sigma$ region. 
Figure~\ref{fig:nonflathubble} shows the distribution of posterior probability of $H_0$ for non-flat $\Lambda$CDM model. As noted earlier the best-fit in this case is slightly higher. Thus, CCHP and SH0ES value are within 1~$\sigma$ region in this case, but the Planck value is just outside 1~$\sigma$ region. It should be noted that in both cases the CCHP value is consistent with the Hubble parameter data. 

\section{Conclusion}
\label{sec:conclusion}
We compare flat and non-flat $\Lambda$CDM cosmologies in the current work and calculate the expansion rate using Hubble parameter data. We have used a variety of statistical techniques to assess the data from differential galaxy ages for this reason. The following are our main conclusions:
(1) Non flat $\Lambda$CDM cosmology favors a higher value of density as well as expansion rate, in comparison to flat $\Lambda$CDM Cosmology. (ii) AIC score is smaller for flat $\Lambda$CDM model which also has less number of parameters. Both the facts make it a better choice. (iii) For the value of Hubble constant, Planck \cite{ade2014planck} and CCHP are consistent with the $\Lambda$CDM results. However SH0ES results are quite high and are not consistent at $1-\sigma$ confidence level. (Iv) SH0ES \cite{riess20162} and CCHP values \cite{freedman2019carnegie} of $H_{0}$ are consistent with our results using non-flat $\Lambda$CDM model as it provides higher value. (v) CCHP value is consistent with Hubble parameter data in both cases as well as with other SNe Ia data \cite{thakur2021cosmological}. (vi) Since, the number of data  points is only 31 and the errror bars in the data are large, the posterior probability curve is wide. A concrete statement about the Hubble tension can be made once we have sufficient number of Hubble parameter data.  

%%%%%%%%%%%%%%%%%%%%%%%%%

\end{document}